%% file: main.tex
\documentclass[conference]{IEEEtran}
\IEEEoverridecommandlockouts
\usepackage{cite}
\usepackage{amsmath,amssymb,amsfonts}
\usepackage{graphicx}
\usepackage{textcomp}
\usepackage{xcolor}
\def\BibTeX{{\rm B\kern-.05em{\sc i\kern-.025em b}\kern-.08em
    T\kern-.1667em\lower.7ex\hbox{E}\kern-.125emX}}

\usepackage{algorithm}
\usepackage{subfigure}
\usepackage[noend]{algpseudocode}
\usepackage{enumitem}
\usepackage{booktabs} 

\begin{document}

\title{Dynamic Graph Collaborative Filtering}



\author{
    \IEEEauthorblockN{Xiaohan Li\IEEEauthorrefmark{1}\IEEEauthorrefmark{2}, Mengqi Zhang\IEEEauthorrefmark{1}\IEEEauthorrefmark{3}\IEEEauthorrefmark{4}, Shu Wu\IEEEauthorrefmark{3}\IEEEauthorrefmark{4}\IEEEauthorrefmark{6}, Zheng Liu\IEEEauthorrefmark{2}, Liang Wang\IEEEauthorrefmark{3}\IEEEauthorrefmark{4}, Philip S. Yu\IEEEauthorrefmark{2}}
    \IEEEauthorblockA{\IEEEauthorrefmark{2}University of Illinois at Chicago, Chicago, IL, USA
    \\\{xli241, zliu212, psyu\}@uic.edu}
    \IEEEauthorblockA{\IEEEauthorrefmark{3}School of Artificial Intelligence, University of Chinese Academy of Sciences, Beijing, China
    \\mengqi.zhang@cripac.ia.ac.cn}
    \IEEEauthorblockA{\IEEEauthorrefmark{4}Institute of Automation, Chinese Academy of Sciences, Beijing, China
    \\\{shu.wu, wangliang\}@nlpr.ia.ac.cn}
    
    \thanks{\IEEEauthorrefmark{1}Both authors contributed equally to this research.}
    \thanks{\IEEEauthorrefmark{6}To whom correspondence should be addressed.}
}

\maketitle

\begin{abstract}
Dynamic recommendation is essential for modern recommender systems to provide real-time predictions based on sequential data. In real-world scenarios, the popularity of items and interests of users change over time. Based on this assumption, many previous works focus on interaction sequences and learn evolutionary embeddings of users and items. However, we argue that sequence-based models are not able to capture collaborative information among users and items directly. Here we propose Dynamic Graph Collaborative Filtering (DGCF), a novel framework leveraging dynamic graphs to capture collaborative and sequential relations of both items and users at the same time. We propose three update mechanisms: zero-order `inheritance', first-order `propagation', and second-order `aggregation', to represent the impact on a user or item when a new interaction occurs. Based on them, we update related user and item embeddings simultaneously when interactions occur in turn, and then use the latest embeddings to make recommendations. Extensive experiments conducted on three public datasets show that DGCF significantly outperforms the state-of-the-art dynamic recommendation methods up to 30\%. Our approach achieves higher performance when the dataset contains less action repetition, indicating the effectiveness of integrating dynamic collaborative information.

 
\end{abstract}





\input{Sections/introduction}
\input{Sections/model}
\input{Sections/experiments}

\input{Sections/related-works}
\input{Sections/conclusions}

\bibliographystyle{IEEEtran}
\bibliography{ref} 

\end{document}

%% file: Sections/introduction.tex
\section{Introduction}
Dynamic recommender systems have proved their effectiveness in many online applications, such as social media, online shopping, and streaming media. They leverage historical interaction sequences of users and items to predict the item that the user may interact with in the future. In real-world scenarios, both user interest and item popularity may shift and evolve along with time. Therefore, it is crucial for a dynamic recommendation model to accurately capture the dynamic changes in user and item perspectives to make accurate predictions. Besides, collaborative information is proved powerful in making recommendation \cite{rendle2010factorizing, koren2009collaborative, rendle2009bpr}. Users that share common interacted items tend to have similar interests, and the methods that leverage this property to recommender system is what we call Collaborative Filtering (CF). Consequently, combining dynamic changes of users and items with collaborative information is one of the main tasks in dynamic recommender systems.

To build a dynamic recommender system, an intuitive way is to model sequences of interactions. Previously, many kinds of sequence-based methods have been developed based on user-item interactions. For example, RNN-based sequential prediction models \cite{hidasi2016session, li2017neural, liu2016context, wu2017recurrent} use the recurrent architecture to capture the long-term dependency of the item sequences. Besides, to take user sequences into consideration as well, Jodie \cite{kumar2019predicting} and other dynamic evolution models \cite{you2019hierarchical, wang2016coevolutionary, wu2019dual} use double RNNs to simultaneously model the update of users and items based on the evolutionary processes of them. We argue that all the models above fail to utilize collaborative information directly. These sequence-based methods primarily model the transition relations between items but ignoring the similarity between users. To mitigate the problem of lacking collaborative information, the graph structure is an alternative instrument for dynamic recommender systems.

\begin{figure}
  \includegraphics[width=0.48\textwidth]{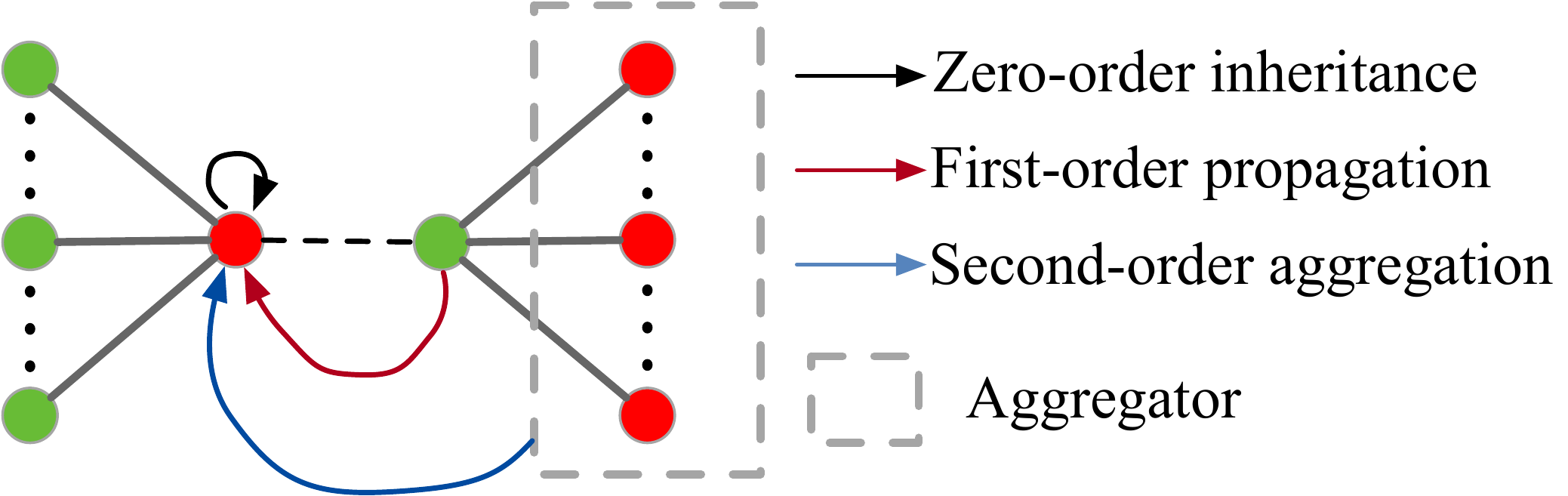}
  \caption{User-Item graph. Red nodes denote users, and green nodes denote items. A solid line means the user has interacted with the item, a dash line means the user and the item are interacting. Arrows in different colors represent different relations when updating the embedding of a node.}
 
  \label{fig:dy_graph_2}
\end{figure}

Several previous works have shown the advantage of using the graph structure in recommender systems, but they all have restrictions in different aspects. According to \cite{wang2019neural, zheng2018spectral}, the graph structure is capable of incorporating collaborative information explicitly. By taking user-item interactions as a bipartite graph, these models exploit the high-order connectivity of users and items and encode collaborative information into the graph structure. However, all these models are only suitable for static scenarios. The advantages of sequential dependency and time information are wasted in them. Moreover, SR-GNN \cite{wu2019session} proves the superiority of graph structures over sequences in the dynamic recommendation, but it fails to incorporate the evolving of items. To deal with these problems, we leverage dynamic graphs to model the evolutionary process of dynamic recommender systems.


In our proposed dynamic graph, nodes are users and items, and edges are their interactions. In the beginning, the graph only contains isolated user and item nodes. With more users, items and their interactions join the dataset, the nodes and edges evolve and grow to a large graph. To model this process and learn the embeddings of users and items at different times, we develop three update mechanisms in the dynamic graph, which are shown in Figure \ref{fig:dy_graph_2}. The first one is \textbf{zero-order inheritance}, in which each node inherits its previous embedding and incorporates the new features of it to update its embedding. Secondly, \textbf{first-order propagation} builds the connection between two sides of the interaction by propagating one's embedding to the other. It updates the embeddings of the user and item in the interaction simultaneously. Finally, \textbf{second-order aggregation} leverages aggregator functions to obtain an overall embedding for all neighbors of the node in the user side, and then pass the embedding to the node in the item side. It is a direct manner to utilize collaborative information.


Based on these three update mechanisms mentioned above, we propose Dynamic Graph Collaborative Filtering (DGCF) to employ all of them under a unified framework. Figure \ref{fig:dyrec_model} illustrates the workflow of the DGCF model. There are three modules in the model, corresponding to the three update mechanisms. Each part produces an embedding, and then the embeddings generated by the three parts are fused to learn the embedding of the node. At the end of our model, we utilize an evolutionary loss to take the time information, i.e., time stamps, into our model to make recommendation. Details of the model will be introduced in section \ref{sec:proposed model}. Overall, our proposed DGCF is capable of directly learning user and item embeddings and perform recommendation tasks in an end-to-end framework.

To summarize, our main contributions are listed as follows:
\begin{itemize}
	\item To the best of our knowledge, our work is the first one to introduce the dynamic graph into dynamic recommendation scenarios to model the interactions and updates between users and items.
	
	\item We design a novel framework for the dynamic recommendation task with  graph structure, which can effectively model the dynamic relationship between users and items.
	
	\item We conduct empirical experiments on three public datasets. Experimental results demonstrate that our DGCF model achieves state-of-the-art results on these datasets, especially for datasets with lower action repetition.
\end{itemize}

%% file: Sections/model.tex
\section{Proposed Model}
\label{sec:proposed model}
In this section, we present the proposed Dynamic Graph Collaborative Filtering (DGCF) in detail. We first formulate the dynamic graph recommendation problem. Then we introduce the embedding update and recommendation modules of the DGCF. Finally, we illustrate the process of optimization and training.


\begin{table}[tb]
  \centering
  \caption{Notations}
  \vspace{2mm}
    \resizebox{0.48\textwidth}{!}{\begin{tabular}{ll}
    \toprule
    Notation                    &Explanation \\
    \midrule
    $t^-, t, t^+$                 &Time points of previous, current,\\
    &and future interaction\\
    $\mathcal{G}_t=(\mathcal{V}_t, \mathcal{E}_t)$                 &Dynamic graph at time $t$\\
    $S_i=(u_i, v_i, t_i, f_i)$       &$i$-th interaction\\
    $u, v$        &User $u$ and item $v$ \\
    $\mathbf{h}_u^t, \mathbf{h}_v^t $ &The embedding of user and item at time $t$ \\
    $\theta, \phi, \zeta$        &Zero, first, and second-order functions \\
    $\mathbf{W}$                      &Weight matrix \\
    $\mathcal{H}_v^u =\{v_1, v_2, ..., v_m\}$       &Second-order neighbors of item $v$ \\
    $\mathcal{H}_u^v =\{u_1, u_2, ..., u_n\}$       &Second-order neighbors of user $u$ \\
    $\operatorname{F}(\cdot )$             &Fusion function \\
    $\operatorname{MLP}$           &Multi-Layer Perceptron\\
    $\mathbf{f}$                    &feature vector \\
    \bottomrule
    \end{tabular}}
  \label{table:notation}
\end{table}

\subsection{Preliminary}

\subsubsection{Dynamic recommendation}
Let $U$, $V$ represent the user and item sets, respectively. In a dynamic recommendation scenario, the $i$-th user-item interaction is represented in a tuple $S_i=(u_i, v_i, t_i, \mathbf{f}_i)$, where $i \in \{1,2,\cdots,I\}$, and $I$ is the total number of interactions. $u_i\in U$, $v_i \in V$ are the user and item in the interaction and $t_i$ is the time stamp. $\mathbf{f}_i$ denotes features of the interaction, and it includes user features $\mathbf{f}_u$ and item features $\mathbf{f}_v$. The target of dynamic recommendation is to learn the representations of the user and item from current interaction and historical records, and then predict the most possible item that the user will interact with in the future.

\begin{figure*}
  \includegraphics[width=1\textwidth]{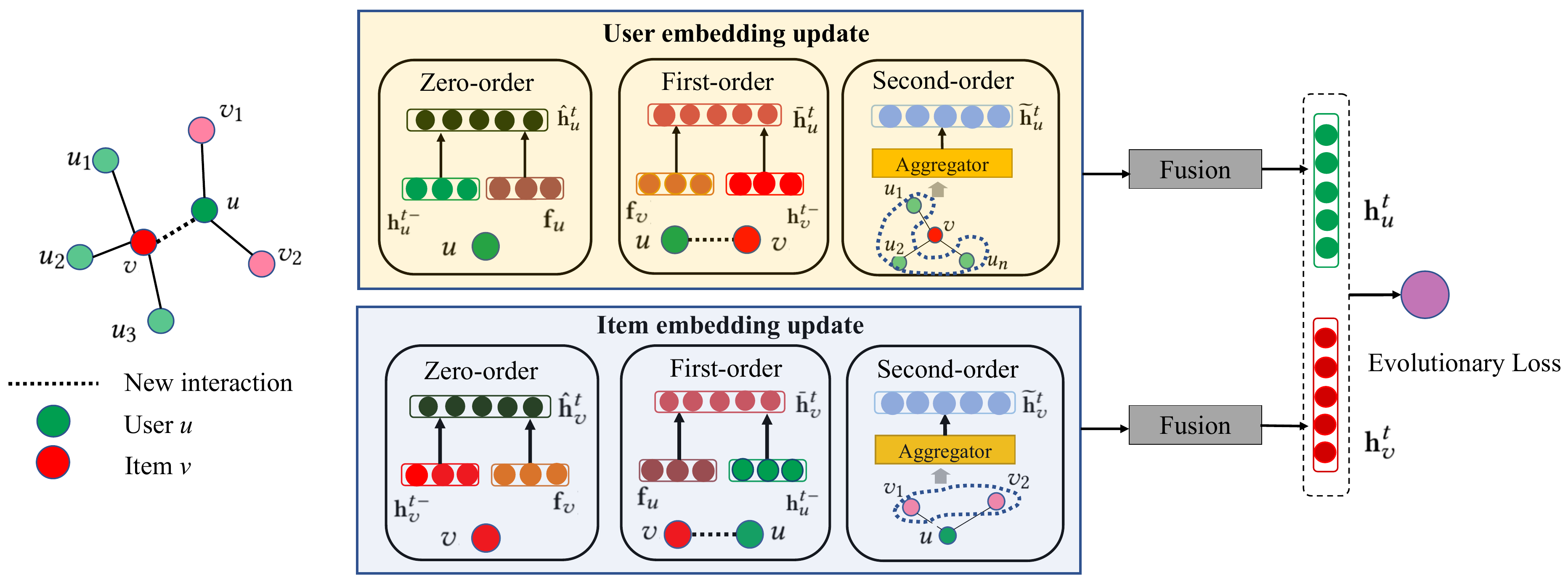}
  \caption{Illustration of the Dynamic Graph Collaborative Filtering (DGCF). Left: A new interaction joins in the user-item graph. Right: Overall structure of DGCF. The pink and light green nodes means the neighbors of the target nodes. $\mathbf{h}_u^{t^-}$ and $\mathbf{h}_v^{t^-}$ denote the embedding of user $u$ and item $v$ before time $t$. Based on three update mechanisms, $\mathbf{h}_u^t$, $\mathbf{h}_v^t$ are produced by our DGCF at the same time. With $\mathbf{h}_u^t$, $\mathbf{h}_v^t$, and evolutionary loss function, we can get the recommendation in the end.}
  \label{fig:dyrec_model}
\end{figure*}

\subsubsection{Dynamic graph}
The interactions between users and items at time stamp $t$ construct a dynamic graph $\mathcal{G}_t=(\mathcal{V}_t, \mathcal{E}_t)$, where $\mathcal{V}_t$, $\mathcal{E}_t$ are the sets of nodes and edges in $\mathcal{G}_t$ respectively. Under recommendation settings, $\mathcal{V}_t$ contains all user and item nodes, and $\mathcal{E}_t$ is the set of all interactions between users and items before time $t$. Essentially, the graph here is a bipartite graph, and all edges are between user and item nodes. We use $\mathbf{h}_u^t \in \mathbb{R}^d$ and $\mathbf{h}_v^t \in \mathbb{R}^d$ to represent the embedding of user node $u$ and item node $v$ at time $t$. The initial graph $\mathcal{G}_{t_0}=(\mathcal{V}_{t_0}, \mathcal{E}_{t_0})$ at time $t_0$ consists of isolated nodes or a snapshot of the dynamic graph, and the initial embeddings of users and items are initial feature vectors or random vectors. Then, while another interaction $S = (u, v, t, \mathbf{f})$ joins the graph, user and item embeddings $\mathbf{h}_u^t$, $\mathbf{h}_v^t$ are updated by our proposed DGCF. Besides, $\mathbf{h}^{t-}_u$ and $\mathbf{h}^{t-}_v$ represent the most recent embeddings of user node $u$ and item node $v$ before time stamp $t$. The notations used throughout this paper are summarized in Table \ref{table:notation}.

\subsection{Overview}
 
Figure \ref{fig:dyrec_model} is an overview of our proposed framework. The interactions between users and items form a dynamic graph over time. When a new user-item interaction joins, we utilize the embedding update mechanism (Section \ref{sec:update}) to update both user and item nodes together. Then, we get the user and item embeddings in the future via the projection functions (Section \ref{sec:recom}). Finally, we calculate the $L2$ distance between the predicted item embedding and all other item embeddings, and then recommend items with the smallest distance to the predicted item embedding.

\subsection{Embedding Update}
\label{sec:update}
First, we discuss the embedding update mechanisms: (1) zero-order `inheritance', which means users and items take the input of their previous embeddings and new features, (2) first-order `propagation', which models current user-item iteration, and (3) second-order `aggregation', which aggregates the previous users that have interacted with the item into the current user, and vice versa. 

\subsubsection{Zero-order `inheritance'}
For a dynamic graph, the node to be updated firstly inherits the influence of the previous state and new features of itself. Similar to existing sequential prediction methods \cite{hidasi2016session,li2017neural,kumar2019predicting}, we use the previous embedding as a part of the input to `inherit' the previous state. Besides, we additionally encode the time interval between current and previous embeddings as a part of the features to learn the user and item embeddings. For user embedding $\mathbf{h}_u$ and item embedding $\mathbf{h}_v$, the forward formulas are
\begin{align}
 &\mathbf{\hat{h}}_u^t = \theta_u(\mathbf{W}_0^u\mathbf{h}_u^{t^-} + \mathbf{w}_0\Delta t_u + \mathbf{W}_0^f\mathbf{f}_u),\\
 &\mathbf{\hat{h}}_v^t = \theta_v(\mathbf{W}_0^v\mathbf{h}_v^{t^-} + \mathbf{w}_0\Delta t_v + \mathbf{W}_0^f\mathbf{f}_v),
\end{align}
where $\Delta t$ is the time interval between current time $t$ and previous interaction time $t^-$, and $\mathbf{h}_u^{t^-}, \mathbf{h}_v^{t^-}$ are the most recent embeddings of user $u$ and item $v$ before $t$.  $\mathbf{f}_u$ and $\mathbf{f}_v$ are the features of the user and the item respectively. $\mathbf{W}_0 \in \mathbb{R}^{d \times d}$ are parameter matrices and $\mathbf{w}_0 \in \mathbb{R}^d$ is the parameter vector to encode time interval $\Delta t$. $\theta_u$ and $\theta_v$ are activation functions. To improve the computational speed, our work here uses an identity map instead of a non-linear activation function.

\subsubsection{First-order `propagation'}

In our model, we build a dynamic bipartite graph to model the interactions between user and item nodes, which means a user node's first-order neighbors are the items that he or she has interacted with, and vice versa. 

In dynamic recommendation scenarios, the item that a user interacts with, to some extent, reflects his or her recent intention and interest. Correspondingly, users who are interested in a specific item can be regarded as a part of the item's properties. Therefore, it is essential to exploit the first-order neighbor information to learn user and item embeddings. Note that our model only incorporates the current interacted node as the input instead of taking all first-order neighbors as inputs.  To be specific, when an interaction involving user $u$ and item $v$ occurs, on the one hand, item $v$'s current embedding and features like reviews or descriptions are incorporated into user $u$'s embedding. On the other hand, the user $u$'s current embedding and features, e.g., the user's profile, are injected into item $v$'s embedding. Formally,
\begin{align}
    &\mathbf{\bar{h}}_u^t = \phi_u(\mathbf{W}_1^u\mathbf{h}_v^{t^-} + \mathbf{W}_1^f\mathbf{f}_v),\\
    &\mathbf{\bar{h}}_v^t = \phi_v(\mathbf{W}_1^v\mathbf{h}_u^{t^-} + \mathbf{W}_1^f\mathbf{f}_u),
\end{align}
where $\mathbf{W}_1 \in \mathbb{R}^{d \times d}$ are parameter matrices. In this way, the features of the current interaction are propagated to update user and item embeddings. Similar to zero-order inheritance, we still use identity map function for $\phi_u$ and $\phi_v$. 

\subsubsection{Second-order `aggregation'}
In the spirit of modeling collaborative filtering in a dynamic graph, the update between nodes considers not only the historical sequence of the node and the information between interacted nodes, but also the structural information among the nodes. The rationale behind second-order aggregation is to model the collaborative relationship between users and items. In a dynamic graph, it captures the influence of the nodes at second-order proximity passing through the other node participating in the interaction.

Specifically, for a user, he or she may have bought a bunch of items before the current interaction, and now this user buys a new item, so we assume the newly purchased item has a collaborative relation with the previously purchased items of the user. To model this relation, we build a direct connection, which denoted as $v_u  \rightarrow u \rightarrow v$. Here $v_u \in \mathcal{H}_v^u$, where $\mathcal{H}_v^u =\{v_1, v_2, ..., v_m\}$ is the set of previously purchased items of user $u$, and $v$ is the item in current interaction. Node $u$ serves as a bridge passing information from $v_u$ to node $v$ so that $v$ receives the aggregated second-order information through $u$. For the user $u$, the relation is $ u_v  \rightarrow v \rightarrow u$, where $u_v \in \mathcal{H}_u^v$, $\mathcal{H}_u^v = \{u_1, u_2, ..., u_n\}$ is the set of users who previously purchased item $v$. 

As shown in Figure \ref{fig:dyrec_model}, to make the second-order information flow from the neighbors of $v$ to $u$, we take $u$ as the anchor node and $u_i \in \mathcal{H}_u^v$ as second-order neighbor nodes in the graph. Then we use aggregator functions to transmit neighborhood information to the anchor node. This process is formulated as
\begin{align}
    &\mathbf{\widetilde{h}}_u^t = \zeta_u(\mathbf{h}_u^{t^-},\mathbf{h}_{u_1}^{t^-}, \mathbf{h}_{u_2}^{t^-},...,\mathbf{h}_{u_n}^{t^-}),\\
    &\mathbf{\widetilde{h}}_v^t = \zeta_v(\mathbf{h}_v^{t^-},\mathbf{h}_{v_1}^{t^-}, \mathbf{h}_{v_2}^{t^-},...,\mathbf{h}_{v_m}^{t^-}),
\end{align}
where $\zeta_u$ and $\zeta_v$ are aggregator functions. In the dynamic graph, users and items are two different kinds of nodes with different properties. For example, the neighbors of a user are the items that he or she has interacted with, which are time-dependent. The neighbors of an item are the users who have interacted with it, which tend to be similar. Besides, the number of users that have interacted with a popular item could be significantly large-scale. Thus, the aggregator functions we use also should be different for user and item nodes. The following part provides some candidate aggregator functions we can use in second-order update:
\begin{itemize}[leftmargin=*]
    \item \textbf{Mean aggregator} is a straightforward operator to aggregate the neighbor information of user $u$ and item $v$. According to \cite{hamilton2017inductive}, \textbf{Mean aggregator} can be viewed as an inductive variant of the GCN \cite{kipf2016semi} approach. So, the formulas of the aggregator can be written as follows:
    \begin{align}
    &\mathbf{\widetilde{h}}_u^t = \mathbf{h}_u^{t^-} + \frac{1}{\lvert \mathcal{H}_v^u \rvert }\sum_{u_i \in {\mathcal{H}_u^v}}\mathbf{W}_u^m \mathbf{h}^{t^-}_{u_i},\\
    &\mathbf{\widetilde{h}}_v^t = \mathbf{h}_v^{t^-} +  \frac{1}{\lvert \mathcal{H}_u^v \rvert }\sum_{v_i \in {\mathcal{H}_v^u}}\mathbf{W}_v^m
    \mathbf{h}^{t^-}_{v_i},
    \end{align}
    where $\mathbf{W}_{\cdot}^m \in \mathbb{R}^{d\times d}$ are aggregation parameters.
    \item \textbf{LSTM aggregator} is a complex aggregator function based on LSTM \cite{hochreiter1997long} architecture. By taking sequential data as inputs, LSTM has strong non-linear memory expression capability to keep track of long term memory. From user's perspective, previous items that users have interacted have explicit sequential dependency, so we feed all of the user node's neighbors into the aggregator function in a chronological order. Besides, from the item's perspective, we order its connected users by time and input them to LSTM. The LSTM aggregator can be formulated as:
    \begin{align}
    &\mathbf{\widetilde{h}}_u^t = \mathbf{h}_u^{t^-} + \operatorname{LSTM}_u(\mathbf{h}_{u_1}^{t^-}, \mathbf{h}_{u_2}^{t^-},...,\mathbf{h}_{u_n}^{t^-}),\\
    &\mathbf{\widetilde{h}}_v^t = \mathbf{h}_v^{t^-} +  \operatorname{LSTM}_v(\mathbf{h}_{v_1}^{t^-}, \mathbf{h}_{v_2}^{t^-},...,\mathbf{h}_{v_m}^{t^-}).
    \end{align}
    
    \item \textbf{Graph Attention aggregator} can compute attention weights between the central node and the neighbor nodes, which indicate the importance of each neighbor node to the central node. Inspired by the GAT \cite{velivckovic2017graph} model, we define graph attention aggregator as:
    \begin{align}
    &\mathbf{\widetilde{h}}_u^t = \sum_{u_i \in{\mathcal{H}_u^v}}\alpha_{ui}\mathbf{h}_{u_i}^{t^-},\\
    &\alpha_{ui} = \frac{\exp(\operatorname{LeakyRelu}(\mathbf{W}_w\lbrack \mathbf{h}_u^{t^-} \parallel \mathbf{h}_{u_i}^{t^-}\rbrack))}{\sum_{u_i \in{\mathcal{H}_u^v}} \exp(\operatorname{LeakyRelu}(\mathbf{W}_w[ h_u^{t^-} \parallel \mathbf{h}_{u_i}^{t^-}]))},
    \end{align}
    \begin{align}
    &\mathbf{\widetilde{h}}_v^t = \sum_{u_i \in{\mathcal{H}_v^u}}\alpha_{vi}\mathbf{h}_{v_i}^{t^-},\\
    &\alpha_{vi} = \frac{\exp(\operatorname{LeakyRelu}(\mathbf{W}_w\lbrack \mathbf{h}_v^{t^-} \parallel \mathbf{h}_{v_i}^{t^-}\rbrack))}{\sum_{v_i \in{\mathcal{H}_v^u}} \exp(\operatorname{LeakyRelu}(\mathbf{W}_w[\mathbf{h}_v^{t^-} \parallel \mathbf{h}_{v_i}^{t^-}]))},
\end{align}
where $\mathbf{W}_w \in \mathbb{R}^{2d}$ is a weight matrix, and $\parallel$ is the concatenation operation.
\end{itemize}

In practical recommendation scenarios, second-order aggregation may face enormous computational costs due to the large scale of data. Therefore, when performing second-order aggregation, we select a fixed number of neighbors for aggregation. We call the number of neighbor nodes selected as aggregator size. 

For higher-order information, we choose not to use it because of the following reasons. Firstly, higher-order information may lead to over-smoothing problem \cite{li2018deeper}, which makes the node embeddings prone to be similar. Besides, using higher-order information increases the computational complexity in power law. As efficiency is a vital issue in recommender system, we try to control the computational complexity to be acceptable in most scenarios and only use up to second-order information.

\subsubsection{Fusion information}
To combine the three kinds of update mechanisms in learning node embeddings in the dynamic graph, we fuse the above mentioned three representations to obtain the final update formula:
\begin{align}
&\mathbf{h}_u^t = \operatorname{F}_u(\mathbf{W}_u^{zero}\mathbf{\hat{h}}_u^t + \mathbf{W}_u^{first}\mathbf{\bar{h}}_u^t + \mathbf{W}_u^{second}\mathbf{\widetilde{h}_u^t} ),\\
&\mathbf{h}_v^t =\operatorname{F}_v(\mathbf{W}_v^{zero}\mathbf{\hat{h}}_v^t + \mathbf{W}_v^{first}\mathbf{\bar{h}}_v^t + \mathbf{W}_v^{second}\mathbf{\widetilde{h}_v^t} ),
\end{align}
where $\mathbf{h}_u^t, \mathbf{h}_v^t \in \mathbb{R}^d$ are the node embeddings updated after the user $u$ interact with item $v$ at time $t$. $\operatorname{F}_u, \operatorname{F}_u$ are fusion functions of user and item respectively. Here we generally choose $\operatorname{sigmod}$ $\sigma(\cdot)$ as activation function. $\mathbf{W}^{zero}$, $\mathbf{W}^{first}$, and $\mathbf{W}^{second} \in \mathbb{R}^{d \times d}$  are parameters to control the influence of three update mechanisms. 

\subsection{Recommendation}
\label{sec:recom}
In dynamic recommendation, the goal is to predict the item that the user is most likely to interact with at time $t$ according to his or her historical interaction sequence before time $t$. Intuitively, this is an analogy to link prediction tasks in dynamic graphs. To be specific, our target is to predict the item node $v$ in the dynamic graph that the user node $u$ is most likely to link to at time $t$. Based on \cite{kumar2019predicting}, we propose an evolutionary loss for dynamic graph.

\subsubsection{Evolution formula}
\label{evol_form}
Different from traditional collaborative filtering methods, our model is designed for predicting future interaction. Specifically, given a future time point, we can leverage our model to predict the future embeddings and then make recommendation. It is a more flexible setting, because the predicted results do not rely on the sequences. Instead, they are based on the embeddings learned by the dynamic graph structure.

Since $\mathbf{h}_u^t$ means the predicted embedding of the user's future, we need an estimated future embeddings to measure whether the predicted embedding is accurate. Motivated by \cite{kumar2019predicting}, we assume the growth of users is smooth, so the embedding vector of the user node evolves in a contiguous space. Therefore, we set a projection function to estimate the future embedding based on element-wise product of the previous embedding and time interval. We define the embedding projection formula of user $u$ after current time $t$ to the future time $t^+$ as follows:
\begin{align}
&\mathbf{\widehat{h}}_u^{t^+} =\operatorname{MLP}_u(\mathbf{h}_u^{t}\odot(\mathbf{1}+\mathbf{w}_t(t^+-t)),
\end{align}
where $\mathbf{w}_t \in \mathbb{R}^d$ is time-context parameter to convert the time interval to vector, $\mathbf{1} \in \mathbb{R}^d$ is a vector with all elements $1$. $MLP$ here means Multi-Layer Perceptron. $t^+$ is the future time that the user interacts with the next item. With this projection function, the future item embedding grows in a smooth trajectory w.r.t. the time interval.

After obtaining the projected embedding $\mathbf{\widehat{h}}_u^{t^+}$ of the user $u$, we learn the future embedding of the item $v$ denoted as $\mathbf{\widehat{h}}_v^{t^+}$ by setting another projection function. The projected item embedding is based on three parts: the user that currently interacts with, the update features of the user and the item itself, which are all we have already known. So, we define the projection formula of item $v$ as:
\begin{align}
&\mathbf{\widehat{h}}_v^{t^+} =\operatorname{MLP}_v(\mathbf{W}_2\mathbf{\widehat{h}}_u^{t^+}+\mathbf{W}_3\mathbf{f}_u + \mathbf{W}_4\mathbf{f}_v)),
\end{align}
where $\mathbf{W}_2$,$\mathbf{W}_3$ and $\mathbf{W}_4$ denote the weight matrix.

\subsubsection{Loss function}

When we have the estimated future embeddings by projection functions, we take them as ground truth embeddings in our loss function. In order to train our model, the loss function is composed of Mean Square Error (MSE) between model-generated embeddings $\mathbf{h}_v^{t}$, $\mathbf{h}_u^{t}$ and estimated ground truth embeddings $\mathbf{\widehat{h}}_v^{t^+}$, $\mathbf{\widehat{h}}_u^{t^+}$ at each interaction time $t$. Besides, we need another constraint for the item embedding to avoid overfitting. We constrain the distance between model-generated $\mathbf{h}_v^{t}$ and mostly recently embedding $\mathbf{h}_v^{t^-}$ of item $v$, and between $\mathbf{h}_u^{t}$ and $\mathbf{h}_u^{t^-}$, respectively,  to make the node embedding more consistent with the previous one. The assumption behind this constraint is that items' and users' properties tend to be stable in a short time. The loss function is written as follows:
\begin{equation}
\begin{split}
\mathcal{L} = \sum_{(u,v,t,f)\in \{S_i\}_{i=0}^I}&\|\mathbf{\widehat{h}}_v^{t^+} - \mathbf{h}_v^{t} \|_2+\lambda_u\|\mathbf{{h}}_u^{t} -  \mathbf{h}_u^{t^-} \|_2+ \\
&\alpha_v\|\mathbf{h}_v^t-\mathbf{h}_v^{t^-}\|_2,
\end{split}
\end{equation}
where $\{S_t\}_{i=0}^I$ denotes the interaction events sorted by chronological order, and $\lambda_u$ and $\alpha_v$ are smooth coefficients, which are used to prevent the embedding of user and item from deviating too much during the update process.


To make recommendations for a user, we calculated the $L2$ distances between the predicted item embedding that we obtain from the loss function and all other item embeddings. Then the nearest Top-$k$ items are what we predict for the user.

Compared with traditional BPR loss \cite{rendle2009bpr}, the evolutionary loss is more suitable for dynamic recommendation, because it takes time into account. As a result, the changing trajectories for users and items are modeled by this loss \cite{kumar2019predicting}, and it can make more precise recommendation for the next item.

\subsection{Optimization and Training}
Similar with Recurrent Neural Networks (RNNs), we apply the back-propagation through time (BPTT) algorithm for model training. The model parameters are optimized by Adam optimizer \cite{kingma2014adam}.

To speed up the training process, we use the same method of constructing batches as \cite{kumar2019predicting}. As mentioned in \cite{kumar2019predicting}, the training algorithm needs to follow two critical criteria: (1) it should process the interactions in each batch simultaneously, and (2) the batching process should maintain the original temporal ordering of the interactions and keep the sequential dependency in the generated embeddings. In practice, we arrange the interaction events $S_i$ in chronological order to get an event sequence \{$S_1, S_2, \cdots, S_{I}\}$ numbered by integer, and $I$ is the total number of interactions. We traverse through the temporally sorted sequence of interactions iteratively and put each interaction to a $Batch_k$, where $k \in [1, I]$. In the initial stage of constructing the $Batch$ sequence: each $Batch$ set is empty at first, and the $Batch$ index is $-1$. We define as $B_{init}(u)=-1$, $B_{init}(v)=-1$. After each interaction $(u,v,t,f_r)$ is added to $Batch$, we update the batch index of the user $u$ and item $v$. For each interaction, the index of the added $Batch$ is $\operatorname{max}\{B(u)+1, B(v)+1)\}$. When the interaction is added to the $Batch$, we update the index of the added $u$ and $v$. This mechanism ensures that the embeddings of users and items in the same $Batch$ are updated simultaneously in the training and testing process.

%% file: Sections/experiments.tex
\section{Experiments}
All source codes and datasets are provided in this link\footnote{https://github.com/CRIPAC-DIG/DGCF}. In this section, we design the experiments to answer the following questions:

\textbf{Q1}: How does DGCF perform compared with other state-of-the-art dynamic or sequential models?

\textbf{Q2}: What is the influence of three types of embedding update mechanisms (zero-order inheritance, first-order propagation, second-order aggregation) in DGCF?

\textbf{Q3}: What is the effect of different aggregator functions in second-order aggregation on model performance?

\textbf{Q4}: How do different hyper-parameters (e.g. aggregation size) affect the performance of DGCF?

\subsection{Datasets Description}
To evaluate the proposed model, we conduct experiments on three real-world datasets. The amounts of users, items and interactions for  datasets and their action repetitions are listed in Table \ref{tab:dataset}. It is worth emphasizing that the three datasets differ significantly in terms of users' repetitive behaviors. Here is the details of the datasets:

{\bf Reddit:} This dataset contains one month of posts made by users on subreddits \cite{baumgartenreddit}. The 10,000 most active users and the 1,000 most active subreddits are selected and treated as users and items respectively, and they have 672,447 interactions. Besides, each post's text is converted into a feature vector to represent their LIWC categories \cite{pennebaker2001linguistic}.

{\bf Wikipedia edits:} This dataset contains one month of edits on Wikipedia\cite{kossinets2012processed}. The editors who made at least 5 edits and the 1,000 most edited pages are filtered out as users and items for recommendation. This dataset contains 157,474 interactions in total. Similarly, the edit texts are converted into an LIWC-feature vector.

{\bf LastFM:} This dataset contains the listening records of users within a month\cite{hidasi2012fast}. 1000 users and the 1000 most listened songs are selected as users and items, and 1,293,103 interactions are in this dataset. Note that interactions do not have features in this dataset.

All user-item interactions are arranged in chronological order. Then we split the training, validation and test set in a proportion of 80\%, 10\%, 10\% for each dataset.
For each interaction $(u, v, t, \mathbf{f})$ in the test set, our goal is to use the given $u$ and $v$ to predict the item that the user is most likely to interact with at time $t$.
\begin{table}[]
    \centering
    \caption{The amount of users, items,  interactions and action repetition rate in each dataset.}
    \resizebox{0.5\textwidth}{!}{\begin{tabular}{ccccc}
    \toprule
         Data & Users & Items & Interactions &Action Repetition  \\
    \midrule
        Reddit & 10000 & 1000 & 672447 &79\% \\
        Wikipedia & 8227 & 1000 & 157474 & 61\% \\
        LastFM & 1000 & 1000 & 1293103 & 8.6\% \\
    \bottomrule
    \end{tabular}}
    
    \label{tab:dataset}
\end{table}

\subsection{Compared Methods}
To evaluate the performance of DGCF, we compare it with the following baseline methods:

\begin{itemize}
    \item \textbf{LSTM} \cite{hochreiter1997long}: It is a variant of RNN whose name is Long Short-Term Memory (LSTM). LSTM updates user (session) embedding by inputting a sequence of historical interacted items of the user into the LSTM cell, which could capture the long-term dependence of the item sequence. 
    
    \item \textbf{Time-LSTM} \cite{zhu2017next}: It uses time gates in LSTM to model time intervals in the interaction sequences.
    
    \item \textbf{RRN} \cite{wu2017recurrent}: Recurrent Recommender Network (RRN) predicts future trajectories to learn user and item embeddings based on LSTM.
    
    \item \textbf{CTDNE} \cite{wang2016coevolutionary}: It is a state-of-the-art model in generating embeddings from temporal networks, but it only produces static embeddings.
    
    \item \textbf{DeepCoevolve} \cite{dai2016deep}: It is based on co-evolutionary point process algorithms. We use 10 negative samples per interaction following the setting of \cite{kumar2019predicting}.
    
    \item \textbf{Jodie} \cite{kumar2019predicting} : It is a state-of-the-art model in dynamic recommendation problem. It defines a projection operation to predict dynamic embedding trajectory.
    
\end{itemize}

\subsection{Experimental Settings}
\subsubsection{\bf Evaluation Metrics}

Two evaluation metrics are used to measure the performance of our DGCF framework: 

(1) \textbf{Mean Reciprocal Rank (MRR)} supposes the model produces a list of recommended items to a user, and the list is ordered by confidence of the prediction. With MRR, we can measure the performance of the model with respect to the ranking list of items. Higher MRR score means target items tend to have higher rank positions in the predicted item lists. Formally, MRR is defined as:

\begin{equation}
    MRR = \frac{1}{|N|} \sum_{i=1}^{|N|}  \frac{1}{rank_u},
\end{equation}
where $rank_u$ represents the rank position of the target item for user $u$.

(2) \textbf{Recall@10} means the number of target items that are in the top-10 recommendation lists. To calculate recall@10, we use equation

\begin{equation}
    Recall@10 = \frac{n_{hit}}{n_{test}},
\end{equation}
where $n_{hit}$ is the number of target items that are among the top-10 recommendation list and $n_{test}$ is the number of all test cases.

\subsubsection{\bf Parameter Settings}
We implement DGCF framework in \emph{Pytorch}\footnote{https://pytorch.org/}. The dimensionality of embeddings is $128$ for all attempts. We use randomly sampled vectors in Gaussian distribution with a mean of 0 and a variance of 1 to initialize embeddings of the users and items. Features of users and items are one-hot vectors. Adam optimizer with learning rate $1e-3$, $L2$ penalty $1e-3$ is adopted in our model. The smoothing coefficients $\lambda$ and $\alpha$ in loss function are set to 1. We run 50 epochs each time and select the best attempt based on the validation set for testing. For comparison methods, we mostly use the default hyperparameters in their paper.

\subsection{Performance Comparison (Q1)}
To prove the superiority of our proposed DGCF model, we conduct experiments on three datasets and compare our model with six baseline methods. Table \ref{table:performance2} shows that DGCF significantly outperforms all six baselines on the three datasets according to both MRR and Recall@10. Especially on LastFM, the improvements are 34.3\% on MRR and 27.7\% on Recall@10. Based on the experimental results shown in Table \ref{table:performance2}, we find three following facts:

\begin{itemize}[leftmargin=*]

    \item  DGCF yields significant improvements on LastFM. Besides, compared with Jodie, the improvements on Reddit, Wikipedia, and LastFM are in an increasing order, which are consistent with the repetitive action pattern in the datasets. The main reason for the improvements might be that DGCF explicitly considers the collaborative information in the graph. LastFM dataset includes users and their listened songs. Intuitively, users tend to listen to different songs that belong to similar genres, and that is why LastFM dataset has a low action repetition. In this case, it is not easy to make recommendations only based on sequences. However, with collaborative information we utilize in DGCF, we can find similar sequences from other users and make recommendations. It can be proof that DGCF can deal with low action repetition situation.
   \item In DGCF, the improvements on MRR are greater than on Recall@10. It may result from projection functions and the evolutionary loss. Because our loss considers the future time, the predicted results corresponding to different time stamps tend to be different. In this case, our model has a better performance on the ranking of recommended items.
\end{itemize}

\begin{table}[t]
  \centering
  \caption{Experiments on three datasets compare DGCF with six baseline models based on Mean Reciprocal Rank (MRR) and Recall@10 (R@10). The bold and underlined numbers mean the best and second-best results on each dataset and metric, respectively. "Improvement" means the minimum improvement among all baselines.}
    \resizebox{0.48\textwidth}{!}{\begin{tabular}{clccccccccc}
    \toprule
    {Models}  
  &\multicolumn{2}{c}{LastFM}  &\multicolumn{2}{c}{Wikipedia}  &\multicolumn{2}{c}{Reddit}  \\
    \cmidrule(lr){2-3}   \cmidrule(lr){4-5}  \cmidrule(lr){6-7}     
    & MRR & R@10 & MRR & R@10 & MRR & R@10  \\
    \midrule 
      LSTM   &  0.081 & 0.127  & 0.332 & 0.459 & 0.367 & 0.573 \\
      Time-LSTM & 0.088 & 0.146  & 0.251 & 0.353 & 0.398 & 0.601   \\
      RRN   &    0.093  & 0.199 & 0.530 & 0.628   & 0.605 & 0.751 \\
      CTDNE  &  0.010   &0.010  &0.035  &0.056   &0.165  &0.257   \\
      DeepCoevolve & 0.021 & 0.042 & 0.515 & 0.563  & 0.243 & 0.305   \\
      Jodie   &   \underline{0.239} & \underline{0.387}  & \underline{0.746} & \underline{0.821}   & \underline{0.724} & \underline{0.851} \\
      DGCF &   \textbf{0.321} & \textbf{0.456}  & \textbf{0.786} & \textbf{0.852} & \textbf{0.726} & \textbf{0.856} \\
     \midrule
     Improvement & 34.3\% & 27.7\% & 5.4\% & 3.6\% & 0.2\% & 0.5\% \\
    \bottomrule
    \end{tabular}}
  \label{table:performance2}
\end{table}

\subsection{Ablation Study (Q2)}
\begin{figure}[tbp]
\centering
\subfigure[MRR]{
\includegraphics[width=2.5in]{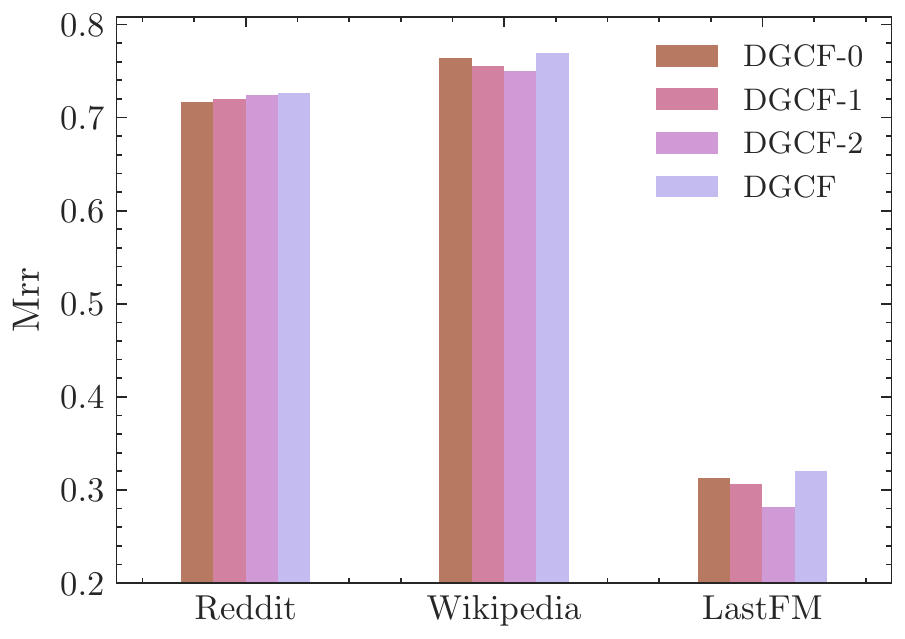}	
}
\quad
\subfigure[Recall@10]{
\includegraphics[width=2.5in]{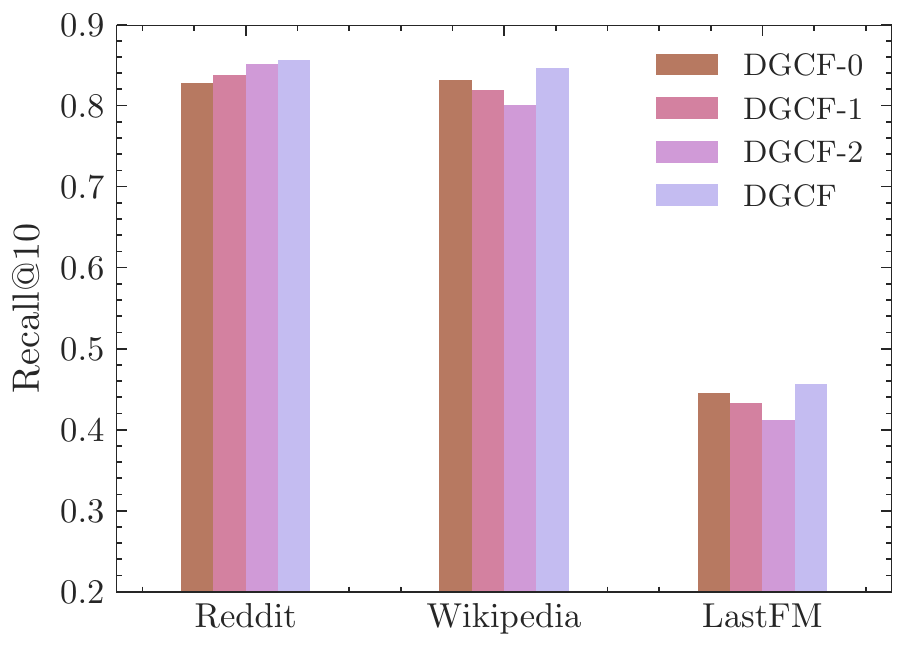}	
}
\caption{Ablation Study. It shows different node relations affect the performance of the model. Due to the different characteristics of datasets, The results are different on several datasets.}
\label{fig:ablation}
\end{figure}

As the three update mechanisms are crucial in DGCF, we measure their effectiveness respectively by conducting experiments. We compare our model with different ablation models: (1) DGCF-0: DGCF without zero-order inheritance. (2) DGCF-1: DGCF without first-order propagation. (3) DGCF-2: DGCF without second-order aggregation. The DGCF here uses graph attention as its aggregator function. Figure \ref{fig:ablation} summarizes the experimental results. 
    
Compared with ablation models, DGCF achieves the best on all datasets, which proves the effectiveness of each module of our model. The following observations are further derived from the results:
\begin{itemize}[leftmargin=*]
\item On the LastFM dataset, we find that DGCF-2 shows evidently lower performance than others. Since the second-order update is capable of finding the users who like the same songs, it is plausible that second-order aggregation is advantageous on the LastFM dataset. For example, when a user does not have many repetitive actions to indicate his/her interests, our model is still capable of modeling the user based on the similarity with other users. As mentioned in Q1, DGCF is able to deal with low action repetition datasets. Without the second-order update, our model cannot achieve similar results on LastFM. As a result, it proves the correctness of this assumption.

\item Compared to DGCF, DGCF-0 and DGCF-1 tend to be stable. It shows that when the model lacks sequential information or current interaction information, second-order aggregation can still have a decent performance by taking advantage of collaborative information.

\end{itemize}

\subsection{Aggregator Function (Q3)}
In this subsection, we test the effectiveness of Mean, LSTM, and Graph attention aggregator functions respectively. The experimental results are shown in Table \ref{table:aggregator}. 

In Table \ref{table:aggregator}, graph attention outperforms the other two aggregator functions on all datasets. It may because that attention mechanism can learn the strength of the relations between the central node and its neighbors. In other words, graph attention is able to explicitly select effective collaborative information among many second-order neighbor nodes to update user or item embeddings. Compared with graph attention, the drawback of mean and LSTM is that neighbors of the central node are equally important when performing aggregation, so that the most influential nodes might be ignored. Different from the other two datasets, low action repetition dataset LastFM depends more on second-order aggregation, so the attention mechanism improves more on it.

\begin{table}[t]
  \centering
  \caption{Aggregator Function. The influence of different aggregator functions on the model performance.}
  \resizebox{0.473\textwidth}{!}{\begin{tabular}{clccccccccc}
    \toprule
    {Aggregator}  
  &\multicolumn{2}{c}{LastFM}  &\multicolumn{2}{c}{Reddit}  &\multicolumn{2}{c}{Wikipedia}  \\
    \cmidrule(lr){2-3}   \cmidrule(lr){4-5}  \cmidrule(lr){6-7}     
    & MRR & R@10 & MRR & R@10 & MRR & R@10  \\
    \midrule 
    Mean & 0.296 & 0.419 & 0.721 & 0.844 & 0.770 & 0.836\\
    LSTM & 0.291 & 0.425 & 0.721 & 0.841 & 0.755 & 0.815\\
    Attention & 0.321 & 0.456 & 0.726 & 0.856 & 0.786 & 0.852\\
   \bottomrule
    \end{tabular}}
  \label{table:aggregator}
\end{table}


\subsection{Hyper-Parameter Study (Q4)}
\begin{figure}[tbp]
\centering
\subfigure[MRR]{
\includegraphics[width=2.5in]{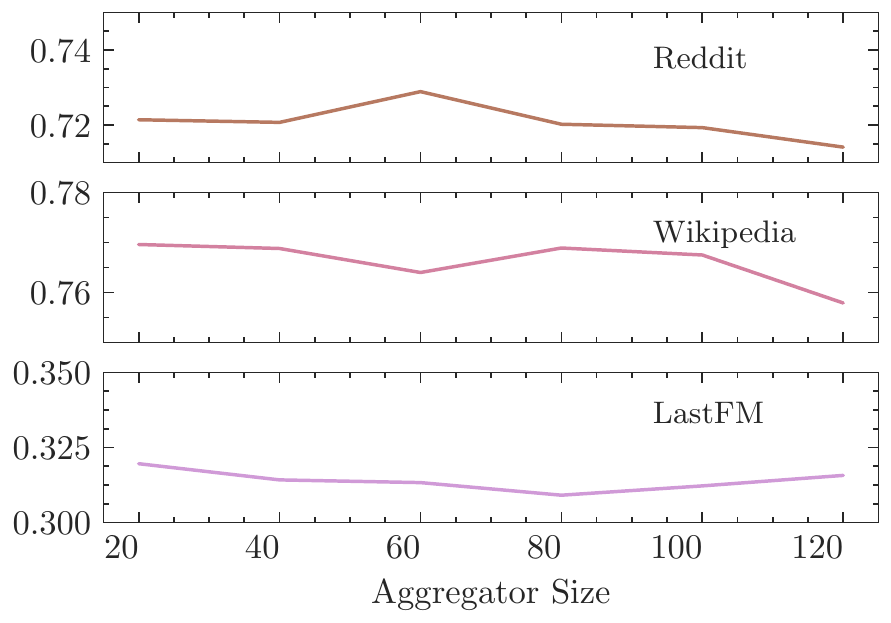}	
}
\quad
\subfigure[Recall@10]{
\includegraphics[width=2.5in]{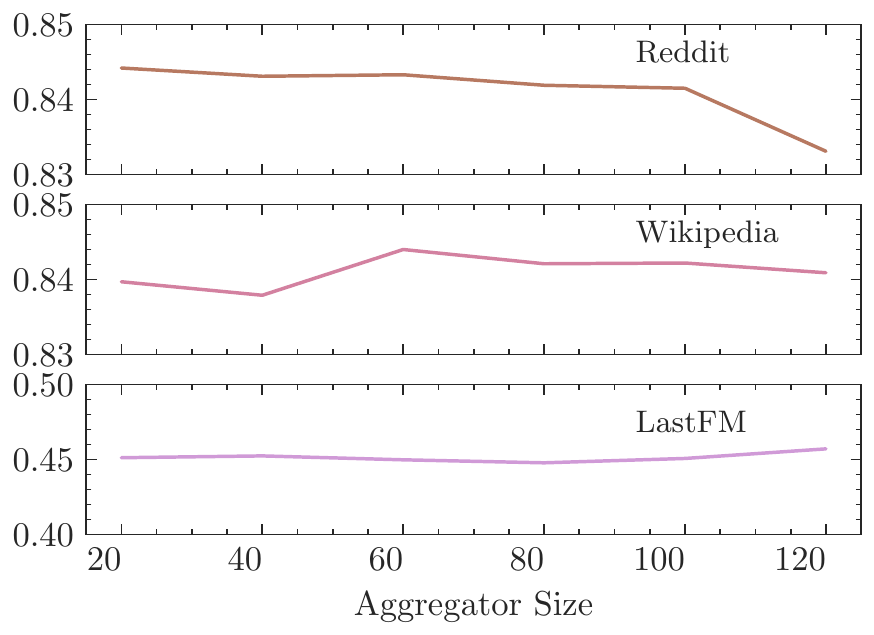}	
}
\caption{Aggregation size. As the aggregator size changes, the performance fluctuation of the model on the datasets is small, indicating that the model is generally robust to aggregator sizes in second-order aggregation.}
\label{fig:size}
\end{figure}

Appropriate aggregator size not only guarantees model performance but also speeds up training and inference. In this section, we test the performance of DGCF when applying different aggregation sizes. We select graph attention aggregator and choose the aggregation size
as ${20, 40, 60, 80, 100, 120}$. Figure \ref{fig:size} shows the results. According to results, a smaller $T$ can lead to higher performance. The degradation of performances when aggregation sizes increase maybe because of the redundancy of second-order collaborative information. Therefore, we can reduce the aggregator size to improve training and inference speed.

\section{Discussion}
In this section, we compare the DGCF with two representative dynamic models, which are RNN and Jodie, to highlight the main differences and innovative parts of our model.

\subsection{Difference between RNN and DGCF}
Recurrent Neural Network (RNN) based models, such as GRU4Rec \cite{hidasi2016session}, NARM \cite{li2017neural}, are widely used in sequential recommendation problems. These models take the user’s historical item sequences as input to model the user’s interest. However, RNN is only a special case of our DGCF when first and second-order relations are removed. It makes RNN fail to consider the variations of items as well as the interactions between users and items.

\subsection{Difference between Jodie and DGCF}
Joint Dynamic User-Item Embeddings (Jodie) \cite{kumar2019predicting} is a representative method of co-evolving based models. It uses two RNNs, which are user-RNN and item-RNN and projection operation to predict the embedding of users and items. Our DGCF model shares similar motivation with Jodie, but compared to DGCF, Jodie is also a special case with no second-order aggregation. This makes it difficult to explicitly model the collaborative relation between users and items. Therefore, when users have few repetitive actions, RNN-based model cannot achieve satisfactory results, and collaborative relations play an important role in this case. This difference results in our significant improvement over Jodie on LastFM dataset, which has the lowest action repetition rate.

%% file: Sections/related-works.tex
\section{Related works}
In this section, we review related works on dynamic recommendation models, graph-based recommendation and dynamic graph representation learning. We also include the comparison between our model and previous methods in the end of each subsection.

\subsection{Dynamic recommendation}
Distinct from static recommendation models like Matrix Factorization (MF) and Bayesian Personal Ranking (BPR), the main task of dynamic recommendation models is to capture the variations of users and items. Recently, Recurrent Neural Network (RNN) and its variants (LSTM and GRU) are widely used in the dynamic recommendation problems. Hidasi et al. \cite{hidasi2016session} apply RNN in session-based recommendation, which is a sub-problem of dynamic recommendation, to model the fluctuation of users' interests. Time-LSTM \cite{zhu2017next} combines LSTM with time gates to model time differences between interactions. Although the RNN model has made great progress in the sequential or dynamic recommendation, it has shortcomings in long-range dependence. To solve the problem of long-distance dependence, NARM, Stamp, and SasRec \cite{li2017neural,liu2018stamp,kang2018self} utilizes attention mechanism to capture users' main purposes, and also improve the training speed. Besides, CNN models \cite{tang2018personalized} are also introduced to dynamic recommendation. However, all these methods only utilize the item trajectory of a user to model variations in the user's interest, while ignoring the evolution of items. 

To deal with this problem, some methods that jointly learn representations of users and items by using the point process model \cite{dai2016deep} and RNN model \cite{wu2017recurrent}. Jodie \cite{kumar2019predicting} predicts user and item embedding with RNN and projection operation. All methods above are recommendation models based on RNNs from different perspectives. They mainly deal with the item sequences or user sequences in the temporal order. However, collaborative signal, which means indirect connections between user-user or item-item, is not used among these works. Therefore, in the DGCF, we not only consider both user and item sequence, but also exploit high-order neighbors in the user-item graph to enrich the training data.

\subsection{Graph-based recommendation}
Users, items, and their interactions can be seen as two types of nodes and edges in a bipartite graph. The advantages of modeling user-item interactions as a graph are: 1) graph-based algorithms like random walk and Graph Neural Networks (GNNs) can be applied to predict links between users and items. 2) High-order connectivity can be explored to enrich training data. Because of the ability to reach high-order neighbors, random walk is tried in making recommendations on the interaction graph. HOP-Rec \cite{yang2018hop} performs random walks on the graph to consider high-order neighbors. RecWalk \cite{nikolakopoulos2019recwalk} is also a random walk-based method that leverages the spectral properties of nearly uncoupled Markov chains for the top-N recommendation. However, random walk-based methods have an issue of lacking robustness.

Nowadays, GNNs show remarkable success in many applications \cite{peng2019fine, dou2020enhancing, li2019fi, zhang2020every, peng2020spatial,Zhu:2020vf,yu2020tagnn}. The effectiveness of GNNs is also proved on recommendation problems. GCMC \cite{berg2017graph} applies Graph Convolutional Networks (GCN) \cite{kipf2016semi} in completing the user-item matrix. PinSage \cite{ying2018graph} introduces GraphSAGE \cite{hamilton2017inductive} into recommender system on item-item graph. Spectral CF \cite{zheng2018spectral} leverage spectral convolution over the user-item bipartite graph to discover possible connections in the spectral domain. SR-GNN \cite{wu2019session} and A-PGNN\cite{wu2019personalizing} use GNNs on session-based recommendation. BasConv \cite{liu2020basconv} leverages graph convolution to user-basket-item graph embedding. NGCF \cite{wang2019neural} propagates user and item embeddings hierarchically to model high-order connectivity. Although these methods achieve significant performances on static recommender system, all of them fail to make use of the influence of time, and the dynamics in users and items are not well considered among these methods. Therefore, we take the graph-based recommendation models under a dynamic graph framework to combine graph structure with time series.

\subsection{Dynamic graph representation learning}
Representation learning over graph-structured data has received wide attention \cite{perozzi2014deepwalk, grover2016node2vec, hamilton2017inductive, tang2015line}. It aims to encode the high-dimensional graph information into low-dimensional vectors. However, in real-world graph data like social networks and citation networks, the graphs are always evolving. To deal with this problem, the graph encoding methods should also consider the dynamics of data, and we call this kind of methods as dynamic graph representation learning.

Previously, some preliminary methods take the evolving graph as snapshots during the discrete-time. DANE \cite{li2017attributed} proposes an embedding method on dynamic attributed network. It models the variations of the adjacency matrix and attribute matrix based on matrix perturbation. DynGEM \cite{goyal2018dyngem} trains a deep autoencoder based model across snapshots of the graph to learn stable graph embeddings over time. TIMERS \cite{zhang2018timers} propose an incremental method based on Singular Value Decomposition (SVD) for dynamic representation learning. DyRep \cite{trivedi2019dyrep} defines association and communication events on dynamic graphs and learning the representation based on graph attention neural networks. Our DGCF model is motivated by these works, especially DyRep, to learn dynamic representations of users and items under recommendation scenarios. However, the recommendation scenario is different from social networks because we have two types of nodes and different influences for dynamic events. Under these circumstances, we develop our DGCF model to capture dynamics on both user and item level. 

%% file: Sections/conclusions.tex
\section{Conclusion}
In this paper, we associate the dynamic graph with the dynamic recommendation scenarios and propose a novel framework based on dynamic graph for dynamic recommendation: Dynamic Graph Collaborative Filtering, abbreviated as DGCF. In DGCF, we design three dynamic node update mechanisms for learning node embedding and making recommendations. Experimental results show that our model outperforms all seven baselines. 

The proposed DGCF is an initial trial of combining dynamic graph with recommender system. Apart from user-item bipartite graph, many other kinds of graph structure can be explored with dynamic graph, e.g., knowledge graph, social network, and attributed graph. 

\section{Acknowledgement}
This work is supported in part by National Key Research and Development Program (2018YFB1402600), and NSF under grants III-1526499, III-1763325, III-1909323, and SaTC-1930941.